  \documentclass[aps,twocolumn,pre,showpacs]{revtex4}

\usepackage{amsmath}

\begin{document}
\title{ Explicit solution of the optimal fluctuation problem \\
for an elastic string in a random medium}
\author{I. V. Kolokolov}
\affiliation{L.D. Landau Institute for Theoretical Physics, Kosygina 2,
Moscow 119334, Russia}
\author{S. E. Korshunov}
\affiliation{L.D. Landau Institute for Theoretical Physics, Kosygina 2,
Moscow 119334, Russia}
\date{September 4, 2009}

\begin{abstract}
The free-energy distribution function of an elastic string in a quenched
random potential, $P_L(F)$, is investigated with the help of the optimal
fluctuation approach. The form of the far-right tail of $P_L(F)$ is found
by constructing the exact solution of the nonlinear saddle-point equations
describing the asymptotic form of the optimal fluctuation.
The solution of the problem is obtained for two different types of boundary
conditions 
and for an arbitrary dimension of the imbedding space \makebox{$1+d$} with
$d$ from the interval \makebox{$0<d<2$}. The results are also applicable
for the description of the far-left tail of the height distribution
function in the stochastic growth problem described by the $d$-dimensional
Kardar-Parisi-Zhang equation.
\end{abstract}



\maketitle

\section{Introduction}

In many physical situations, the behavior of some extended manifold is
determined by the competition between its internal elasticity and
interaction with external random potential, which at all reasonable time
scales can be treated as quenched. A number of examples of such a kind
includes domain walls in magnetic materials, vortices and vortex lattices
in type-II superconductors, dislocations and phase boundaries in crystals,
as well as some types of biological objects. It is expected that the
presence of a quenched disorder makes at least some aspects of the
behavior of such systems analogous to those of other systems with quenched
disorder, in particular, spin glasses.

Following Ref. \onlinecite{KZ}, in the case when internal dimension of an
elastic manifold is equal to 1 (that is, an object interacting with a
random potential is an elastic string), the systems of such a kind are
traditionally discussed under the generic name of a directed polymer in a
random medium. In such a case the problem turns out to be formally
equivalent \cite{HHF} to the problem of a stochastic growth described by
the Kardar-Parisi-Zhang (KPZ) equation \cite{KPZ}, which in its turn can
be reduced to the Burgers equation \cite{Burgers} with random force (see
Refs. \onlinecite{Kardar-R} and \onlinecite{HHZ} for reviews).

The investigation of $P_L(F)$, the free-energy distribution function for a
directed polymer (of a large length $L$) in a random potential, was
initiated by Kardar \cite{Kardar}, who proposed an asymptotically exact
method for the calculation of the moments $Z_n\equiv \overline{Z^n}$ of
the distribution of the partition function $Z$ in a $(1+1)\,$-dimensional
system (a string confined to a plane) with a $\delta$-correlated random
potential and made an attempt of expressing the moments of $P_L(F)$ in
terms of $Z_n$. Although soon after that, Medina and Kardar \cite{MK} (see
also Refs. \onlinecite{Kardar-R} and \onlinecite{DIGKB}) realized that the
implementation of the latter task is impossible, the knowledge of $Z_n$
allowed Zhang \cite{Zhang} to find the form of the tail of $P_L(F)$ at
large negative $F$. The two attempts of generalizing the approach of Ref.
\onlinecite{Zhang} to other dimensions were undertaken by Zhang
\cite{Zhang90} and Kolomeisky \cite{Kolom}.

Quite recently, it was understood \cite{KK07,KK08} that the method of Ref.
\onlinecite{Zhang} allows one to study only the most distant part of the
tail (the far-left tail), where $P_L(F)$ is not obliged to have the
universal form \makebox{$P_L(F)=P_*(F/F_*)/F_*$} (with $F_*\propto
L^\omega$) it is supposed to achieve in the thermodynamic limit,
$L\rightarrow\infty$. For $(1+1)\,$-dimensional systems the full form of
the universal distribution function is known from the ingenious exact
solution of the polynuclear growth (PNG) model by Pr\"{a}hofer and Spohn
\cite{PS}. However, there is hardly any hope of generalizing this approach
to other dimensions or to other forms of random potential distribution.

One more essential step in the investigation of different regimes in the
behavior of $P_L(F)$ in systems of different dimensions has been made
recently \cite{KK07,KK08} on the basis of the optimal fluctuation
approach. The original version of this method was introduced in the 1960s
for the investigation of the deepest part of the tail of the density of
states of quantum particles localized in a quenched random potential
\cite{HL,ZL,L67}. Its generalization to Burgers problem has been
constructed in Refs. \onlinecite{GM} and \onlinecite{BFKL}, but for the
quantities which in terms of the directed polymer problem are of no direct
interest, in contrast to the distribution function $P_L(F)$ studied in
Refs. \onlinecite{KK07} and \onlinecite{KK08}. Another accomplishment of
Refs. \onlinecite{KK07} and \onlinecite{KK08} consists in extending the
optimal fluctuation approach to the region of the universal behavior of
$P_L(F)$, where the form of this distribution function is determined by an
effective action with scale-dependent renormalized parameters and does not
depend on how the system is described at microscopic scales.

In the current work, the results of Refs. \onlinecite{KK07} and
\onlinecite{KK08} describing the behavior of $P_L(F)$ at the largest
positive fluctuations of the free energy $F$ (where they are not described
by the universal distribution function) are rederived at a much more
quantitative level by explicitly finding the form of the optimal
fluctuation which is achieved in the limit of large $F$. This allows us
not only to verify the conjectures used earlier 
for finding the scaling behavior of $S(F)\equiv -\ln[P_L(F)]$ in the
corresponding regime, but also to establish the exact value of the
numerical coefficient entering the expression for $S(F)$. For brevity, we
call the part of the right tail of $P_L(F)$ studied below the far-right
tail. The outlook of the paper is as follows.

In Sec. \ref{II}, we formulate the continuous model which is traditionally
used for the quantitative description of a directed polymer in a random
medium and remind how it is related to the KPZ and Burgers problems. Sec.
\ref{OFA} briefly describes the saddle-point problem which has to be
solved for finding the form of the most optimal fluctuation of a random
potential leading to a given value of $F$. In Sec. \ref{ExSol}, we
construct the exact solution of the saddle-point equations introduced in
Sec. \ref{OFA} for the case when the displacement of a considered elastic
string is restricted to a plane (or, in other terms, the transverse
dimension of the system $d$ is equal to 1). We do this for sufficiently
large positive fluctuations of $F$, when the form of the solution becomes
basically independent on temperature $T$ and therefore can be found by
setting $T$ to zero.

However, the solution constructed in Sec. \ref{ExSol} turns out to be not
compatible with the required boundary conditions. Sec. \ref{FreeIC} is
devoted to describing how this solution has to be modified to become
compatible with free initial condition and in Sec. \ref{FixedIC}, the same
problem is solved for fixed initial condition. In both cases, we find the
asymptotically exact (including a numerical coefficient) expression for
$S(F)$ for the limit of large $F$. In Sec. \ref{d>1}, the results of the
two previous sections are generalized for the case of an arbitrary $d$
from the interval \makebox{$0<d<2$}, whereas the concluding Sec.
\ref{concl} is devoted to summarizing the results.

\section{Model \label{II}}

In the main part of this work, our attention is focused on an elastic
string whose motion is confined to a plane. The coordinate along the
average direction of the string is denoted $t$ for the reasons which will
become evident few lines below and $x$ is string's displacement in the
perpendicular direction. Such a string can be described by the Hamiltonian
\begin{equation}                                          \label{H}
H\{x(t)\}
=\int_{0}^{L}dt \left\{\frac{J}{2}\left[\frac{d{ x}(t)}{dt}\right]^2
+V[t,{ x}(t)]\right\} \;,
\end{equation}
where the first term describes the elastic energy and the second one the
interaction with a random potential $V(t,x)$, with $L$ being the total
length of a string along axis $t$. Note that the form of the first term in
Eq. (\ref{H}) relies on the smallness of the angle between the string and
its preferred direction.

The partition function of a string which starts at $t=0$ and ends
at the point $(t,{x})$ is then given by the functional integral
which has exactly the same form as the Euclidean functional integral
describing the motion of a quantum-mechanical particle (whose mass is given
\makebox{by $J$)} in a time-dependent random potential $V(t,{ x})$ (with
$t$ playing the role of imaginary time and temperature $T$ of Plank's
constant $\hbar$).
As a consequence, the evolution of this partition function 
with the increase in $t$ is governed \cite{HHF}
by the imaginary-time Schr\"{o}dinger equation
\begin{equation}                                        \label{dz/dt}
-T{\dot z}
=\left[-\frac{T^2}{2J}\nabla^2+V(t,{ x})\right]z(t,{ x})\,.
\end{equation}
Here and below, a dot denotes differentiation with respect to $t$ and
$\nabla$  differentiation with respect to $x$.

Naturally, $z(t,{ x})$ depends also on the initial condition at $t=0$.
In particular, fixed initial condition, \makebox{${ x}(t=0)={ x}_0$},
corresponds to $z(0,{ x})=\delta({ x}-{ x}_0)$,
whereas free initial condition (which implies the absence
of any restrictions  \cite{KK08,DIGKB} on $x$ at $t=0$) to
\begin{equation}                                            \label{FBC}
z(0,{ x})=1\,.
\end{equation}
Below, the solution of the problem is found for both these types of
initial condition.

It follows from Eq. (\ref{dz/dt}) that the evolution of
the free energy corresponding to $z(t,{ x})$,
\begin{equation}                                             \label{}
f(t,{ x})=-T\ln\left[z(t,{ x})\right]\,,
\end{equation}
with the increase in $t$ is governed \cite{HHF} by the KPZ equation
\cite{KPZ}
\begin{equation}                                             \label{KPZ}
{\dot f}+\frac{1}{2J}(\nabla f)^2-\nu \nabla^2 f = V(t,{ x})\,,
\end{equation}
with the inverted sign of $f$, where $t$ plays the role of time and
$\nu\equiv{T}/{2J}$ of viscosity. On the other hand, the derivation of Eq.
(\ref{KPZ}) with respect to ${ x}$ allows one to establish the equivalence
\cite{HHF} between the directed polymer problem and Burgers equation
\cite{Burgers} with random potential force
\begin{equation}                                             \label{Burg}
{\dot { u}}+u\nabla { u}
-\nu\nabla^2{ u}={J}^{-1}\nabla V(t,{ x})\,,
\end{equation}
where $u(t,x)=\nabla f(t,x)/J$ plays the role of velocity. Note that in
terms of the KPZ problem, the free initial condition (\ref{FBC})
corresponds to starting the growth from a flat interface, \makebox{$f(0,{
x})=\mbox{const}$,} and in terms of the Burgers problem, to starting the
evolution from a liquid at rest, ${ u}(0,{ x})=0$.

To simplify an analytical treatment, the statistic of a random potential
$V(t,{ x})$ is usually assumed to be Gaussian with
\begin{equation}                                              \label{VV}
\overline{V(t,{ x})}=0\,,~~~
\overline{V(t,{ x})V(t',{ x}')}=U(t-t',{ x}-{ x}')\,,
\end{equation}
where an overbar denotes the average with respect to disorder.
{Although the analysis below is focused exclusively on the case of purely
$\delta$-functional correlations,
\begin{equation}                                              \label{U(t,x)}
    U(t-t',{ x}-{ x}')=U_0\delta(t-t')\delta(x-x')\,,
\end{equation}
the results we obtain are applicable also in situations when the
correlations of $V(t,x)$ are characterized by a finite correlation radius
$\xi$ because in the considered regime, the characteristic size of the
optimal fluctuation grows with the increase in $L$ and therefore for
large-enough $L$, the finiteness of $\xi$ is of no importance and an
expression for $U(t-t',{ x}-{ x}')$ can be safely replaced by the
right-hand side of Eq. (\ref{U(t,x)}) with
\begin{equation}                                              \label{}
    U_0=\int_{-\infty}^{+\infty}dt\int_{-\infty}^{+\infty}dx\, U(t,{ x})\;.
\end{equation}

\section{Optimal fluctuation approach\label{OFA}}

We want to find probability of a large positive fluctuation of free energy
of a string which at $t=L$ is fixed at some point $x=x_L$. It is clear
that in the case of free initial condition, the result cannot depend on
$x_L$, so for the simplification of notation, we assume below $x_L=0$ and
analyze fluctuations of $F=f(L,0)-f(0,0)$. As in other situations
\cite{HL,ZL,L67}, the probability of a sufficiently large fluctuation of
$F$ is determined by the most probable fluctuation of a random potential
$V(t,x)$ among those leading to the given value of $F$.

In its turn, the most probable fluctuation of $V(t,x)$ can be found
\cite{FKLM} by looking for the extremum of the Martin-Siggia-Rose action
\cite{MSR,dD,Jans} corresponding to the KPZ problem,
\begin{eqnarray} S\{f,{V}\} & = &
\frac{1}{U_0}\int\limits_0^L \! dt\int_{-\infty}^{+\infty}\! dx
\left\{-\frac{1}{2}{V}^2+ \right.                          \nonumber\\
&& +\left.
{V}\left[\dot f+\frac{1}{2J}\left(\nabla f\right)^2-
\nu\nabla^2f\right]
\right\}\;,                                                 \label{S(f,V)}
\end{eqnarray}
both with respect to $f\equiv f(t,x)$ and to a random potential realization
$V\equiv V(t,x)$.
The form of Eq. (\ref{S(f,V)}) ensures that its variation with
respect to $V(t,x)$ reproduces the KPZ equation (\ref{KPZ}),
whose substitution back into Eq. (\ref{S(f,V)}) reduces it to the
expression
\begin{equation}                                             \label{S(V)}
S\{V\}=\frac{1}{2U_0}\int_{0}^{L}dt \int_{-\infty}^{+\infty}\!\!dx\,
V^2(t,x)\,.
\end{equation}
determining the probability of a given realization of a random potential,
\makebox{${\cal P}\{V\}\propto\exp(-S\{V\})$.} On the other hand,
variation of Eq. (\ref{S(f,V)}) with respect to $f(t,x)$ shows that the
time evolution of the optimal fluctuation of a random potential is
governed by equation \cite{Fogedb}
\begin{eqnarray}
\dot {V}+\frac{1}{J}\nabla\left({V}\nabla f\right)+\nu\nabla^2{V} &
=& 0\,,                                                 \label{peur-mu}
\end{eqnarray}
whose form  implies that the integral of ${V}(t,x)$
over $dx$ is a conserved quantity.

Our aim consist in finding the solution of
Eqs. (\ref{KPZ}) and (\ref{peur-mu})
satisfying condition
\begin{equation}                                         \label{gusl-f}
f(L,0)-f(0,0)=F\,,
\end{equation}
as well as an appropriate initial condition at $t=0$. The application of
this procedure corresponds to calculating the full functional integral
determining $P_L(F)$ with the help of the saddle-point approximation. In
the framework of this approximation, the condition (\ref{gusl-f}) [which
formally can be imposed by including into the functional integral
determining $P_L(F)$ the corresponding $\delta$-functional factor] leads
to the appearance of the condition on $V(t,x)$ at $t=L$ \cite{FKLM},
\begin{equation}                                         \label{gusl-mu}
{V}(L,x)=\lambda\delta(x)\,,
\end{equation}
where, however, the value of $\lambda$ should be chosen to ensure
the fulfillment of condition (\ref{gusl-f}).

The conditions for the applicability of the saddle-point approximation
for the analysis of the far-right tail of $P_L(F)$ are given by $S\gg 1$
and $F\gg JU_0^2L/T^4$. The origin of the former inequality is evident,
whereas the fulfillment of the latter one ensures the possibility
to neglect the renormalization of the parameters of the
system by small-scale fluctuations \cite{cond}.
We also assume that $F \gg T$, which ensures that the characteristic length
scale of the optimal fluctuation is sufficiently large to neglect the
presence of viscous terms in Eqs. (\ref{KPZ}) and (\ref{peur-mu})
\cite{cond}.
This allows us to replace Eqs. (\ref{KPZ}) and (\ref{peur-mu}) by
\begin{subequations}                                   \label{df&dmu/dt}
\begin{eqnarray}
\dot f+\frac{1}{2J}\left(\nabla f\right)^2 & = & {V},\quad
                                                      \label{df/dt}\\
\dot {V}+\frac{1}{J}\nabla\left({V}\nabla f\right) & = & 0\,,
\label{dmu/dt}
\end{eqnarray}
\end{subequations}
which formally corresponds to considering the original (directed polymer)
problem at zero temperature, $T=0$, where the free energy of a string is
reduced to its ground state energy. In accordance with that, in the $T=0$
limit $f(L,0)$ is given by the minimum of Hamiltonian (\ref{H}) on all
string's configurations $x(t)$ which at $t=0$ satisfy a chosen initial
condition and at $t=L$ end up at $x(L)=0$.

Exactly like Eq. (\ref{peur-mu}), Eq. (\ref{dmu/dt}) implies that ${V}(t,x)$
behaves itself like a density of a conserved quantity, but takes into
account only the nondissipative component to the flow of ${V}$ given by
${V} u$, where
\begin{equation}                                 \label{}
u\equiv u(t,{ x})\equiv{J}^{-1}\nabla f(t,{ x})
\end{equation}
plays the role of velocity. Naturally, for $\nu=0$ the time evolution of
$u$ is governed by the nondissipative version of the force-driven Burgers
equation (\ref{Burg}),
\begin{equation}                                     \label{du/dt}
    \dot u+u\nabla u=\nabla{V}/J\;.
\end{equation}

\section{Exact solution of the saddle-point equations \label{ExSol}}

It is clear from symmetry that in the optimal fluctuation we are looking
for, both $f(t,x)$ and ${V}(t,x)$ have to be even functions of $x$. After
expanding them at $x=0$ in Taylor series, it is easy to verify that an
exact solution of Eqs. (\ref{df&dmu/dt}) can be constructed by keeping in
each of these expansions only the first two terms:
\begin{subequations}                                     \label{f&mu}
\begin{eqnarray}
f(t,x) & = & J\left[A(t)-B(t)x^2\right]\,,  \label{f}   \\
{V}(t,x) & = & J\left[C(t)-D(t)x^2\right]\,. \label{mu}
\end{eqnarray}
\end{subequations}
Substitution of Eqs. (\ref{f&mu}) into Eqs. (\ref{df&dmu/dt})
gives then a closed system of four equations,
\begin{subequations}                                      \label{a&b}
\begin{eqnarray}
\dot{A} & = & C\,, \\
\dot{B} & = & 2B^2+D\,,                                   \label{eq-b}
\end{eqnarray}
\end{subequations}
\vspace*{-7mm}
\begin{subequations}                                      \label{c&d}
\begin{eqnarray}
\dot{C} & = & 2BC\,,~~~~~ \\
\dot{D} & = & 6BD\,,                                      \label{eq-d}
\end{eqnarray}
\end{subequations}
which determines the evolution of coefficients $A$, $B$, $C$ and $D$ with
the increase in $t$.

It is easy to see that with the help of Eq. (\ref{eq-d}), $D(t)$ can be
expressed in terms of $B(t)$, which allows one to transform Eq.
(\ref{eq-b}) into a closed equation for $B(t)$,
\begin{equation}                                         \label{eqn-b}
   \dot B=2B^2+D(t_0)\exp\left[6\int_{t_0}^{t}dt'B(t')\right]\;.
\end{equation}
After making a replacement
\begin{equation}                                           \label{b}
 B(t)=-\frac{\dot{\varphi}}{2\varphi}\;,
\end{equation}
Eq. (\ref{eqn-b}) is reduced  to an equation of the Newton's type,
\begin{equation}                                           \label{Neut}
\ddot{\phi}+\frac{\alpha}{\phi^2}=0\;,
\end{equation}
where $\alpha\equiv 2D(t)\phi^3(t)$ is an integral of motion
which does not depend on $t$.

Eq. (\ref{Neut}) can be easily integrated which allows one to ascertain
that its general solution can be written as
\makebox{$\phi(t)=\phi_0\Phi[(t-t_0)/L_*]$},
where $t_0$ and $\phi_0\equiv\phi(t_0)$ are arbitrary constants,
\begin{equation}                                            \label{L*}
    L_*=\frac{\pi}{4[D(t_0)]^{1/2}}\,
\end{equation}
plays the role of the characteristic time scale, and $\Phi(\eta)$ is an
even function of its argument implicitly defined in the interval
$-1\leq\eta\leq 1$ by equation
\begin{equation}                                          \label{Phi}
\sqrt{\Phi(1-\Phi)}+\arccos\sqrt{\Phi}=\frac{\pi}{2}|\eta|\;.
\end{equation}
With the increase of $|\eta|$ from $0$ to $1$, $\Phi(\eta)$ monotonically
decreases from $1$ to $0$. In particular, on approaching
\makebox{$\eta=\pm 1$}, the behavior of $\Phi(\eta)$ is given by
\begin{equation}                                          \label{Phi(1)}
\Phi(\eta)\approx[(3\pi/4)(1-|\eta|)]^{2/3}\,.
\end{equation}
Since it is clear from the form of Eq. (\ref{b}) that the constant
$\phi_0$ drops out from the expression for $B(t)$,
one without the loss of generality can set $\phi_0=1$ and
\begin{equation}                                         \label{phi(t)}
\phi(t)=\Phi\left(\frac{t-t_0}{L_*}\right)\,.
\end{equation}
The functions $A(t)$, $B(t)$, $C(t)$ and $D(t)$
can be then expressed in terms of $\phi\equiv\phi(t)$ as
\begin{subequations}                                    \label{a-d}
\begin{eqnarray}
A(t) & = & A_0+\mbox{sign}(t-t_0)\left.\frac{C_0}{D_0^{1/2}}\right.
\,\arccos\sqrt{\phi}\;,           \label{a(t)}\\
                                                        \label{b(t)}
B(t) & = & 
\mbox{sign}(t-t_0)\left[D_0(1-\phi)/{\phi^3}\right]^{1/2}\;, \\
C(t) & = & {C_0}/{\phi}\;,                              \label{c(t)}\\
D(t) & = & {D_0}/{\phi^3}\;,                            \label{d(t)}
\end{eqnarray}
\end{subequations}
where $A_0=A(t_0)$, $C_0=C(t_0)$ and $D_0=D(t_0)$.

Thus we have found an exact solution of Eqs. (\ref{df&dmu/dt}) in which
$f(t,x)$ is maximal at $x=0$ (for $t>t_0$) and the value of $f(t,0)$
monotonically grows with the increase in $t$. However, the optimal
fluctuation also have to satisfy particular boundary conditions. The
modifications of the solution (\ref{phi(t)})-(\ref{a-d}) compatible with
two different types of initial conditions, free and fixed, are constructed
in Secs. \ref{FreeIC} and \ref{FixedIC}, respectively.

\section{Free initial condition \label{FreeIC}}

When the initial end point of a polymer (at $t=0$) is not fixed (that is,
is free to fluctuate), the boundary condition at $t=0$ can be written as
$z(0,x)=1$ or
\[
f(0,x)=0\,.
\]
Apparently, this condition is compatible with Eq. (\ref{f}) and
in terms of functions $A(t)$ and $B(t)$ corresponds to
\begin{equation}                                           \label{ab=0}
A(0)=0\,,~~B(0)=0\,,
\end{equation}
from where $\dot\phi(0)=0$ and $t_0=0$. However, it is clear that the
solution described by Eqs. (\ref{phi(t)})-(\ref{ab=0}) cannot be the
optimal one because it does not respect condition (\ref{gusl-mu}) which
has to be fulfilled at $t=L$. Moreover, this solution corresponds to an
infinite action and the divergence of the action is coming from the
regions where potential ${V}(t,x)$ is negative, which evidently cannot be
helpful for the creation of a large positive fluctuation of $f(t,0)$.

From the form of Eqs. (\ref{H}) and (\ref{S(V)}), it is clear that
any region where \makebox{$V(t,x)<0$} cannot increase the 
energy of a string 
but makes a positive contribution to the action. Therefore, in a really
optimal fluctuation with $F>0$, potential $V(t,x)$ should be either
positive or zero. In particular, since just the elastic energy of any
configuration $x(t)$ which somewhere crosses or touches the line
\begin{equation}                                                \label{}
x_*(t)=\left[{2F(L-t)}/{J}\right]^{1/2}
\end{equation}
and at $t=L$ ends up at $x(L)=0$ is already larger than $F$,
there is absolutely no reason for $V(t,x)$ to be nonzero
at least for $|x|>x_{\rm *}(t)$.

It turns out that the exact solution of the saddle-point equations
(\ref{df&dmu/dt}) in which potential $V(t,x)$ satisfies
boundary condition (\ref{gusl-mu}) and constraint $V(t,x)\geq 0$
can be constructed on the basis of the solution found in Sec. \ref{ExSol}
just by cutting the dependences (\ref{f&mu}) at the points
\begin{equation}                                       \label{x*}
x=\pm X_{}(t)\,,~~
X_{}(t)\equiv\left[\frac{C(t)}{D(t)}\right]^{1/2}\hspace*{-4mm}
=\left(\frac{C_0}{D_0}\right)^{1/2}\hspace*{-3mm}\phi(t)\,,
\end{equation}
where ${V}(t,x)$ is equal to zero, and replacing them at
\makebox{$|x|>X_{}(t)$} by a more trivial solution of the same equations
with \makebox{${V}(t,x)\equiv 0$} which at $x=\pm X_{}(t)$ has the same
values of $f(t,x)$ and $u(t,x)$ as the solution at $|x|\leq X_{}$. Such a
replacement can be done because the flow of ${V}$ through the moving point
$x=X_{}(t)$ in both solutions is equal to zero. In accordance with that,
the integral of ${V}(t,x)$ over the interval $-X_{}(t)<x<X_{}(t)$ does not
depend on $t$. It is clear from Eq. (\ref{x*}) that $X_{}(t)$ is maximal
at $t=t_0$ and at $t>t_0$ monotonically decreases with the increase of
$t$.

The form of $f(t,x)$ at $|x|>X_{}(t)$ is then given by
\begin{equation}                                      \label{f1}
    f(t,x)=f[t,X_{}(t)]+J\int_{X_{}(t)}^{|x|}dx'\,u_0(t,x')\,,
\end{equation}
where $u_0(t,x)$ is the solution
of Eq. (\ref{du/dt}) with zero right-hand side in the region $x>X(t)$
which at $x=X(t)$ satisfies boundary condition
\begin{equation}                                      \label{BounCond}
    u_0[t,X_{}(t)]=v(t)\,.
\end{equation}
In Eq. (\ref{BounCond}), we have taken into account that in the solution
constructed in Sec. \ref{ExSol} \makebox{$u[t,X_{}(t)]= -2B(t)X_{}(t)$}
coincides with
\begin{equation}                                      \label{v}
v(t) = \frac{dX_{}}{dt}
=\left(\frac{C_0}{D_0}\right)^{1/2}\hspace*{-1.3mm}\dot\phi
=-2\sqrt{C_0(\phi^{-1}-1)}\;,
\end{equation}
the velocity of the point $x=X_{}(t)$. This immediately follows from Eq.
(\ref{b}) and ensures that the points where spacial derivatives of
$u(t,x)$ and ${V}(t,x)$ have jumps always coincide with each other. It is
clear from Eq. (\ref{v}) that $v(t_0)=0$, whereas at $t>t_0$, the absolute
value of $v(t)<0$ monotonically grows with the increase in $t$.

Since Eq. (\ref{du/dt}) with vanishing right-hand side implies that
the velocity of any Lagrangian particle does not depend on time, its
solution satisfying boundary condition (\ref{BounCond}) can be written as
\begin{equation}                                      \label{u0}
u_0(t,x)= v[\tau(t,x)]\,,
\end{equation}
where function $\tau(t,x)$ is implicitly defined by equation
\begin{equation}                                      \label{tau}
x = X_{}(\tau)+(t-\tau)v(\tau)\,.
\end{equation}
Monotonic decrease of $v(\tau)<0$ with the increase in $\tau$ ensures that
in the interval \makebox{$X_{}(t)<x<X_{0}\equiv X(t_0)$}, Eq. (\ref{tau})
has a well-defined and unique solution which at fixed $t$ monotonically
decreases from $t$ at $x=X_{}(t)$ to $0$ at $x=X_{0}$. In accordance with
that, $u_0(t,x)$ as a function of $x$ monotonically increases from
$v(t)<0$ at $x=X_{}(t)$ to $0$ at $x=X_{0}$. For free initial condition
(implying $t_0=0$), the form of the solution at $x>X_0$ remains the same
as in the absence of optimal fluctuation, that is, $u_0[t,x>X_0]\equiv 0$.
The fulfillment of the inequality $\partial u_0(t,x)/\partial x\leq 0$ in
the interval $x>X(t)$ demonstrates the absence of any reasons for the
formation of additional singularities (such as shocks), which confirms the
validity of our assumption that the form of the solution can be understood
without taking into account viscous terms in saddle-point equations
(\ref{KPZ}) and (\ref{peur-mu}).

Substitution of Eqs. (\ref{u0}) and (\ref{tau}) into Eq. (\ref{f1}) and
application of Eqs. (\ref{df/dt}) and (\ref{du/dt}) allow one to reduce
Eq. (\ref{f1}) to
\begin{equation}                                      \label{f1-b}
    f(t,x)=\frac{J}{2}\int\limits_{0}^{\tau(t,x)}d\tau'(t-\tau')
    \frac{dv^2(\tau')}{d\tau'}\,,
\end{equation}
from where it is immediately clear that on approaching $x=X_{0}$, where
$\tau(t,x)$ tends to zero, $f(t,x)$ also tends to zero, so that at
$|x|>X_{0}$, the free energy $f(t,x)$ is equal to zero (that is, remains
exactly the same as in the absence of optimal fluctuation). However, for
our purposes, the exact form of the solution at $|x|>X_{}(t)$ is of no
particular importance, because this region does not contribute anything to
the action.

It is clear that the compatibility of the constructed solution with
condition (\ref{gusl-mu}) is achieved when the interval
$[-X_{}(t),X_{}(t)]$ where the potential is non-vanishing shrinks to a
point. This happens when the argument of function $\Phi$ in Eq.
(\ref{phi(t)}) is equal to 1, that is when
\begin{equation}                                             \label{t0+L*}
    t_0+L_*=L\,,
\end{equation}
which for $t_0=0$ corresponds to $L_*=L$ and
\begin{equation}                                      \label{d0}
    D_0=\left(\frac{\pi}{4L}\right)^2.
\end{equation}
On the other hand, Eq. (\ref{a(t)}) with $A_0=0$ gives
\makebox{$A(L)=(\pi/2) C_0/D_0^{1/2}=2LC_0$}. With the help of the
condition $A(L)=F/J$, following from Eq. (\ref{gusl-f}) this allows one to
conclude that
\begin{equation}                                       \label{c0}
    C_0=\frac{F}{2JL}\;.
\end{equation}

Thus, for free initial condition, the half width of the region where the
optimal fluctuation of a random potential is localized is equal to
\begin{equation}                                        \label{}
X_{0}=\left(\frac{C_0}{D_0}\right)^{1/2}
     =\frac{2}{\pi}\left(\frac{2FL}{J}\right)^{1/2}
\end{equation}
at $t=0$ (when it is maximal) and monotonically decreases to zero as
$X(t)=X_0\Phi(t/L)$ when $t$ increases  to $L$. On the other hand,
${V}(t,0)$, the amplitude of the potential, is minimal at $t=0$ (when it
is equal to $F/2L$) and monotonically increases to infinity. In the
beginning of this section, we have argued that $V(t,x)$ has to vanish at
least for \makebox{$|x|>x_*(t)=[2F(L-t)/J]^{1/2}$} and indeed it can be
checked that $X(t)<x_*(t)$ at all $t$, the maximum of the ratio
$X(t)/x_*(t)$ being approximately equal to $0.765$.

In the case of free initial condition, the optimal fluctuation of a random
potential at $T=0$ has to ensure that $E(x_0)$, the minimum of $H\{x(t)\}$
for all string's configurations with $x(0)=x_0$ and $x(L)=0$, for all
values of $x_0$ should be equal or larger than $F$. In particular, for any
$x_0$ from the interval $|x_0|<X_0$ where the potential is nonzero, the
corresponding energy $E(x_0)$ has to be exactly {equal} to $F$, otherwise
there would exist a possibility to locally decrease the potential without
violating the condition $E(x_0)\geq F$.

The configuration of a string, $x(t)$, which minimizes $H\{x(t)\}$
in the given realization of a random potential for the given values
of $x_0$ and $x(L)$, at $0<x<L$ has to satisfy equation
\begin{equation}                                            \label{extrem}
    -J\frac{d^2x}{dt^2}+\frac{\partial V(t,x)}{\partial x}=0\,,
\end{equation}
which is obtained by the variation of Hamiltonian (\ref{H}) with respect
to $x$. It is not hard to check that for the optimal fluctuation found
above, the solution of this equation for an arbitrary $x_0$ from the
interval \makebox{$-X_0<x_0<X_0$} can be written as
\begin{equation}                                            \label{x(t)}
    x(t)=\frac{x_0}{X_0}X(t)\,.
\end{equation}
All these solutions have the same energy, $E(x_0)=F$.

The value of the action corresponding to the optimal fluctuation can be
then found by substituting Eqs. (\ref{mu}), (\ref{c(t)}) and (\ref{d(t)})
into the functional (\ref{S(V)}), where the integration over $dx$
should be restricted to the interval $-X(t)<x<X(t)$, which gives
\begin{equation}                                       \label{def-b}
S_{\rm free}=\frac{8}{15}\frac{C_0^{5/2}}{D_0^{1/2}}\frac{J^2}{U_0}
\int\limits_0^{L}\frac{dt}{\phi(t)}=
\frac{4\pi}{15}\frac{C_0^{5/2}}{D_0}\frac{J^2}{U_0}\;.
\end{equation}
The integral over $dt$ in Eq. (\ref{def-b}) can be calculated
with the help of replacement \makebox{$dt/\phi=d\phi/(\phi\dot{\phi})$}
and is equal to $\pi/2D_0^{1/2}$.
Substitution of relations (\ref{d0}) and (\ref{c0}) allows one to rewrite
Eq. (\ref{def-b}) in terms of the parameters of the original system as
\begin{equation}                                    \label{Sfree}
S_{\rm free}(F,L)=K\frac{F^{5/2}}{U_0J^{1/2}L^{1/2}}\;,~~~
K=\frac{8\sqrt{2}}{15\pi}\;.
\end{equation}

The exponents entering Eq. (\ref{Sfree}) have been earlier found in Ref.
\onlinecite{KK07} from the scaling arguments based on the assumption that
for large $L$, the form of the optimal fluctuation involves a single
relevant characteristic length scale with the dimension of $x$ which
algebraically depends on the parameters of the system (including $L$) and
grows with the increase of $L$ \cite{exp}. The analysis of this section
has explicitly confirmed this assumption and has allowed us to find the
exact value of the numerical coefficient $K$.

Since the characteristic length scale of the solution we constructed is
given by \makebox{$X_{0}\equiv X(t_0)\sim(FL/J)^{1/2}$}, the neglect of
viscosity $\nu$ remains justified as long as the characteristic relaxation
time corresponding to this length scale
$\tau_{\rm rel}\sim X_{0}^2/\nu\sim FL/T$
is much larger than the time scale of this solution $L$, which
corresponds to
\begin{equation}                                               \label{F>T}
    F \gg T\,.
\end{equation}
Another condition for the validity of Eq. (\ref{Sfree}) is the condition
for the direct applicability of the optimal fluctuation approach. One can
disregard any renormalization effects as long as the characteristic
velocity inside optimal fluctuation is much larger \cite{KK08} than the
characteristic velocity of equilibrium thermal fluctuations at the length
scale $x_c\sim T^3/JU_0$, the only characteristic length scale with the
dimension of $x$ which exists in the problem with
\makebox{$\delta$-functional} correlations (that is, can be constructed
from $T$, $J$ and $U_0$). In terms of $F$, this condition reads
\begin{equation}                                       \label{NoRenorm}
    F\gg U_0^2JL/T^4\,.
\end{equation}
It is easy to check that the fulfillment of conditions (\ref{F>T}) and
(\ref{NoRenorm}) automatically ensures $S\gg 1$, which also is a necessary
condition for the applicability of the saddle-point approximation.

For $L\gg L_{c}$, where $L_{c}\sim T^5/JU_0^2$ is the only characteristic
length scale with the dimension of $L$ which exists in the problem with
$\delta$-functional correlations, condition (\ref{F>T}) automatically
follows from condition (\ref{NoRenorm}) which can be rewritten as $F\gg
(L/L_{c})T$. Thus, for a sufficiently long string (with $L\gg L_{c}$), the
only relevant restriction on $F$ is given by Eq. (\ref{NoRenorm}).

\section{Fixed initial condition \label{FixedIC}}

When both end points of a string are fixed \makebox{[$x(0)=x_0$,
$x(L)=x_L$],} one without the loss of generality can consider the problem
with $x_0=x_L$. Due to the existence of so-called tilting symmetry
\cite{SVBO}, the only difference between the problems with $x_0=x_L$ and
\makebox{$x_0\neq x_L$} consists in the shift of the argument of $P_L(F)$
by \makebox{$\Delta F\equiv J(x_L-x_0)^2/2L$}. For this reason, we
consider below only the case $x_0=x_L=0$.

When a string is fastened at $t=0$ to the point $x=0$, in terms of
$z(t,x)$, the boundary condition at $t=0$ can be written as
$z(0,x)\propto\delta(x)$. In such a case, the behavior of $f(t,x)$ at
$t\rightarrow 0$ is dominated by the elastic contribution to energy, which
allows one to formulate the boundary condition in terms of $f(t,x)$ as
\cite{KK08}
\begin{equation}                                           \label{FixedBC}
    \lim_{t\rightarrow 0}\left[f(t,x)-f^{(0)}(t,x)\right]=0\,,
\end{equation}
where $f^{(0)}(t,x)=Jx^2/2t$ is the free energy of the same system in the
absence of a disorder. Since we are explicitly analyzing only the
$T\rightarrow 0$ limit, we omit the linear in $T$ contribution to the
expression for $f^{(0)}(t,x)$ which vanishes in this limit.
The fulfillment of condition (\ref{FixedBC}) can be ensured,
in particular, by setting
\begin{equation}                                         \label{f(eps)}
    f(\varepsilon,x)=f^{(0)}(\varepsilon,x)=Jx^2/2\varepsilon\,,
\end{equation}
which corresponds to suppressing the noise in the interval $0<t<\varepsilon$,
and afterwards taking the limit $\varepsilon\rightarrow 0$.
Naturally, the free initial condition also can be written
in the form (\ref{FixedBC}) but with $f^{(0)}(t,x)=0$.

Quite remarkably, initial condition (\ref{f(eps)}) is compatible with the
structure of the solution constructed in Sec. \ref{ExSol} and in terms
of functions $A(t)$ and $B(t)$ corresponds to
\begin{equation}                                          \label{a&b(eps)}
    A(\varepsilon)=0,~~~ B(\varepsilon)=-1/2\varepsilon\,.
\end{equation}
Substitution of Eqs. (\ref{L*}), (\ref{Phi(1)}) and (\ref{phi(t)}) into
Eq. (\ref{b(t)}) allows one to establish that for $\varepsilon\ll L_*$,
the condition $B(\varepsilon)=-1/2\varepsilon$ corresponds to
\begin{equation}                                             \label{t0-L*}
    t_0-L_*\approx \varepsilon/3\,.
\end{equation}

Exactly like in the case of free initial condition (see Sec.
\ref{FreeIC}), we have to assume that at \makebox{$|x|>X_{}(t)$} the
dependences (\ref{f}) and (\ref{mu}) are replaced, respectively, by Eq.
(\ref{f1}) and \makebox{${V}(t,x)=0$}. The compatibility with condition
(\ref{gusl-mu}) is achieved then when the interval
\makebox{$-X_{}(t)<x<X_{}(t)$} where the potential is nonvanishing shrinks
to a point, the condition for which is given by Eq. (\ref{t0+L*}). A
comparison of Eq. (\ref{t0+L*}) with Eq. (\ref{t0-L*}) allows one to
conclude that for initial condition (\ref{a&b(eps)}) $L_*\approx
L/2-\varepsilon/6$ and $t_0\approx L/2+\varepsilon/6$, which after taking
the limit $\varepsilon\rightarrow 0$ gives
\begin{equation}                                             \label{L*t0}
    L_*=L/2\,,~~~t_0=L/2\,.
\end{equation}
This unambiguously defines the form of the solution for the case of
fixed initial condition.

In this solution, the configuration of ${V}(t,x)$ is fully symmetric not
only with respect to the change of the sign of $x$ but also with respect
to replacement
\begin{equation}                                            \label{t->}
t \Rightarrow L-t\,.
\end{equation}
The origin of this property is quite clear. In terms of an elastic string,
the problem we are analyzing now  is fully symmetric with respect to
replacement (\ref{t->}), therefore it is quite natural that the spacial
distribution of the potential in the optimal fluctuation also has to have
this symmetry.

Since we are considering the limit of zero temperature when the free
energy of a string is reduced to its energy, which in its turn is just the
sum of the energies of the two halves of the string, the form of the
potential $V(t,x)$ in the symmetric optimal fluctuation can be found
separately for each of the two halves after imposing free boundary
condition at $t=L/2$. This form can be described by Eqs. (\ref{phi(t)}),
(\ref{c(t)}) and (\ref{d(t)}) with $L_*=t_0=L/2$, where the values of
$C_0$ and $D_0$ can be obtained from Eqs. (\ref{c0}) and (\ref{d0}),
respectively, by replacement
\begin{equation}                                            \label{FL}
    F\Rightarrow F/2\,,~~~L\Rightarrow L/2\,.
\end{equation}

The value of the action corresponding to the optimal fluctuation can be
then found by making the same replacement 
in Eq. (\ref{Sfree}) and multiplying the result by the factor of 2,
\begin{equation}                                           \label{Sfix}
    S_{\rm fix}(F,L)=2S_{\rm free}(F/2,L/2)
    =\frac{1}{2}S_{\rm free}(F,L)\,.
\end{equation}
Naturally, the conditions for the applicability of Eq. (\ref{Sfix}) are
the same as for Eq. (\ref{Sfree}) (see the two last paragraphs of
\makebox{Sec. \ref{FreeIC}}). The claim that the optimal fluctuation is
symmetric with respect to replacement (\ref{t->}) and therefore both
halves of the string make equal contributions to its energy can be
additionally confirmed by noting that the sum
\begin{equation}                                               
    S_{\rm free}(F',L/2)+S_{\rm free}(F-F',L/2)
\end{equation}
is minimal when $F'=F-F'=F/2$.

Like in the case of free initial condition, the form of the optimal
fluctuation is such that the whole family of extremal string's
configurations satisfying Eq. (\ref{extrem}) is characterized by the same
value of energy, $E\equiv H\{x(t)\}=F$. Formally, this family again can be
described by  Eq. (\ref{x(t)}) where $x_0$ now should be understood not as
$x(0)$ but more generally as $x(t_0)$.

\section{Generalization to other dimensionalities \label{d>1}}

The same approach can be applied in the situation when polymer's
displacement is not a scalar quantity but a $d$-dimensional vector ${\bf
x}$. In such a case, the expressions for the action and for the
saddle-point equations retain their form, where now operator $\nabla$
should be understood as vector gradient. A spherically symmetric solution
of Eqs. (\ref{df&dmu/dt}) can be then again found in the form (\ref{f&mu})
with $x^2\equiv {\bf x}^2$.

For arbitrary $d$ substitution of Eqs. (\ref{f&mu}) into Eqs.
(\ref{df&dmu/dt}) reproduces Eqs. (\ref{a&b}) in exactly the same form,
whereas Eqs. (\ref{c&d}) are replaced by
\begin{subequations}                                     \label{c&d'}
\begin{eqnarray}
    \dot{C} & = & 2dBC\,, \\
    \dot{D} & = & (4+2d)BD\,.
\end{eqnarray}
\end{subequations}
A general solution of Eqs. (\ref{a&b}) and (\ref{c&d'}) can be then
written as
\begin{subequations}                                    \label{a-d'}
\begin{eqnarray}
A(t) & = & A_0+\mbox{sign}(t-t_0)\frac{C_0I_{-}(\phi,d)}{2(dD_0)^{1/2}}
\;,           \label{a(t)'}\\
                                                        \label{b(t)'}
B(t) & = & 
\mbox{sign}(t-t_0)\left[\frac{D_0}{d}\frac{1-\phi^{d}}{\phi^{2+d}}
\right]^{1/2}, \\
C(t) & = & {C_0}/{\phi^d}\;,                              \label{c(t)'}\\
D(t) & = & {D_0}/{\phi^{2+d}}\;,                            \label{d(t)'}
\end{eqnarray}
\end{subequations}
where
\begin{equation}                                        \label{}
    \phi\equiv\phi(t)=\Phi\left(\frac{t-t_0}{L_*}\right)\,
\end{equation}
with
\begin{equation}                                          \label{}
    L_*=\frac{I_{+}(0,d)}{2(dD_0)^{1/2}}\;
\end{equation}
and $\Phi(\eta)$ is an even function of its argument implicitly defined in
the interval $-1\leq\eta\leq 1$ by equation
\begin{equation}                                          \label{Phi'}
I_{+}(\Phi,d))=I_{+}(0,d)|\eta|\;.
\end{equation}
Here, $I_{\pm}(\phi,d)$ stands for the integral
\begin{equation}                                           \label{Ipm}
I_{\pm}(\phi,d)=
\int_{\phi^{d}}^{1}dq\,\frac{q^{1/d-1\pm 1/2}}{(1-q)^{1/2}}\,,
\end{equation}
in accordance with which $I_{\pm}(0,d)$ is given by the Euler beta
function
\begin{equation}                                           \label{}
I_{\pm}(0,d)=
B\left(\frac{1}{2}\;,\frac{1}{d}\pm\frac{1}{2}\right)=
\frac{\Gamma\left(\frac{1}{2}\right)\Gamma\left(\frac{1}{d}\pm\frac{1}{2}
\right)}
{\Gamma\left(\frac{1}{d}+\frac{1}{2}\pm\frac{1}{2}\right)}\;.
\end{equation}
From the form of Eqs. (\ref{Phi'}) and (\ref{Ipm}), it is clear that with
the increase of $|\eta|$ from 0 to 1, the function $\Phi(\eta)$
monotonically decreases from 1 to 0. It is not hard to check that at
$d=1$, Eqs. (\ref{a-d'}) and (\ref{Phi'}) are reduced to Eqs. (\ref{a-d})
and (\ref{Phi}), respectively.

Exactly like in the case $d=1$, for free initial condition one gets
$t_0=0$, $L_*=L$, $A_0=0$ and $A(L)=F/J$, from where
\begin{equation}                                           \label{C0D0'}
    C_0=\frac{2-d}{2}\frac{F}{JL}\,,~~~
    D_0=\frac{1}{d}\left[\frac{I_+(0,d)}{2L}\right]^2\;.
\end{equation}
On the other hand, Eq. (\ref{def-b}) is replaced by
\begin{equation}                                           \label{Sfree'}
S_{\rm free}=\frac{4\Omega_d}{d(d+2)(d+4)}
\frac{C_0^{1+d/2}}{D_0^{d/2}}\frac{JF}{U_0}\;,
\end{equation}
where $\Omega_d=2\pi^{d/2}/\Gamma(d/2)$ is the area of a $d$-dimensional
sphere. Substitution of Eqs. (\ref{C0D0'}) into Eq. (\ref{Sfree'}) then gives
\begin{subequations}                                      \label{Sfree''}
\begin{equation}                                          \label{Sfree''a}
S_{\rm free}(F,L)= K_d \frac{F^{2+d/2}}{U_0J^{d/2}L^{1-d/2}}\;.
\end{equation}
with
\begin{equation}                                            \label{K}
    K_d=\frac{8(2-d)^{1+d/2}(2d)^{d/2-1}}{(d+2)(d+4)\Gamma(d/2)}
    \left[\frac{\Gamma(1/d+1)}{\Gamma(1/d+1/2)}\right]^d.
\end{equation}
\end{subequations}
Naturally, at $d=1$ numerical coefficient $K_d$ coincides with coefficient
$K$ in Eq. (\ref{Sfree}). Like in the case of \makebox{$d=1$}, the value
of the action for fixed initial condition, \makebox{${\bf x}(t=0)=0$}, can
be found by making in Eq. (\ref{Sfree''a})  replacement (\ref{FL}) and
multiplying the result by the factor of 2, which gives
\begin{equation}                                           \label{Sfixed'}
S_{\rm fix}(F,L)=2S_{\rm free}\left({F}/{2},{L}/{2}\right)
    =\left(\frac{1}{2}\right)^{\,d}S_{\rm free}(F,L)\,.
\end{equation}

The exponents entering Eq. (\ref{Sfree''a}) and determining the dependence
of $S_{\rm free}$ on the parameters of the system have been earlier found
in Ref. \onlinecite{KK08} from the scaling arguments based on the
assumption that for large $L$, the form of the optimal fluctuation
involves a single relevant characteristic length scale with the dimension
of $\bf x$ which algebraically depends on the parameters of the system
(including $L$) and grows with the increase of $L$ \cite{exp}. However,
the analysis of this section reveals that this length scale,
$X_0=(C_0/D_0)^{1/2}$, tends to zero when $d$ approaches $2$ from below,
as well as the value of the action given by Eqs. (\ref{Sfree''}).

This provides one more evidence that at $d\geq 2$, the problem with purely
delta-functional correlations of a random potential becomes ill defined
\cite{ill-reg} and has to be regularized in some way, for example, by
introducing a finite correlation length for the random potential
correlator. In such a situation, the geometrical size of the optimal
fluctuation is determined by this correlation length \cite{KK08} and its
shape is not universal, that is, depends on the particular form of the
random potential correlator. Thus, the range of the applicability of Eqs.
(\ref{Sfree''})  is restricted to $0<d<2$ and includes only one physical
dimension, $d=1$.

\section{Conclusion \label{concl}}

In the current work, we have investigated the form of $P_L(F)$, the
distribution function of the free energy of an elastic string with length
$L$ subject to the action of a random potential with a Gaussian
distribution. This has been done in the framework of the continuous model
traditionally used for the description of such systems, Eq. (\ref{H}). Our
attention has been focused on the far-right tail of $P_L(F)$, that is on
the probability of a very large positive fluctuation of free energy $F$ in
the regime when this probability is determined by the probability of the
most optimal fluctuation of a random potential leading to the given value
of $F$.

We have constructed the exact solution of the nonlinear saddle-point
equations describing the asymptotic form of the optimal fluctuation in the
limit of large $F$ when this form becomes independent of temperature. This
has allowed us to find not only the scaling from of
\makebox{$S(F)=-\ln[P_L(F)]$} but also the value of the numerical
coefficient in the asymptotic expression for $S(F)$.

The solution of the problem has been obtained for two different types of
boundary conditions (corresponding to fixing either one or both end points
of a string) and for an arbitrary dimension of the imbedding space
\makebox{$1+d$} with $d$ from the interval $0<d<2$ ($d$ being the
dimension of the displacement vector). Quite remarkably, in both cases the
asymptotic expressions for $S(F)$, Eqs. (\ref{Sfree''}) and
(\ref{Sfixed'}), are rather universal. In addition to being independent of
temperature, they are applicable not only in the case of
$\delta$-correlated random potential explicitly studied in this work, but
also (for a sufficiently large $L$) in the case of potential whose
correlations are characterized by a finite correlation radius. Note that
our results cannot be compared to those of Brunet and Derrida \cite{BD}
because these authors have considered a very specific regime when the
transverse size of a system (with cylindrical geometry) scales in a
particular way with its length $L$.

Due to the existence of the equivalence \cite{HHF} between the directed
polymer and KPZ problems, the distribution function of the directed
polymer problem in situation when only one of the end points is fixed (and
the other is free to fluctuate) describes also the fluctuations of height
\cite{KK07,KK08} in the \makebox{$d$}-dimensional KPZ problem in the
regime of nonstationary growth which have started from a flat
configuration of the interface, $L$ being the total time of the growth.
The only difference is that the far-right tail of $P_L(F)$ studied in this
work in the traditional notation of the KPZ problem \cite{KPZ} corresponds
to the far-left tail of the height distribution function. In terms of the
KPZ problem, the independence of the results on temperature is translated
into their independence on viscosity.
\begin{center}
{\bf Acknowledgments}
\end{center}
The authors are grateful to G. Blatter and V. B. Geshkenbein for useful
discussions. This work has been supported by the RFBR Grant No.
09-02-01192-a and by the RF President Grants for Scientific Schools No.
4930.2008.2 (I.V.K) and No. 5786.2008.2 (S.E.K.).


\begin{thebibliography}{99}


\bibitem{KZ}      M. Kardar and Y.-C. Zhang,
                  Phys. Rev. Lett. {\bf 58}, 2087 (1987).
\bibitem{HHF}     D. A. Huse, C. L. Henley and D. S. Fisher,
                  Phys. Rev. Lett. {\bf 55}, 2924 (1985).
\bibitem{KPZ}     M. Kardar, G. Parisi and Y.-C. Zhang,
                  Phys. Rev. Lett. {\bf 56}, 889 (1986).
\bibitem{Burgers} J. M. Burgers, {\em The Nonlinear Diffusion Equation}
                  (Reidel, Boston, 1974).
\bibitem{Kardar-R}M. Kardar, in {\em Fluctuating Geometries in Statistical
            Mechanics and Field Theory}, Lecture Notes at Les Houches,
            1994, edited by F. David, P. Ginsparg and
            J. Zinn-Justin (Elsevier Science, Amsterdam, 1996),
            also available as arXiv:cond-mat/9411022.
\bibitem{HHZ}     T. Halpin-Healy and Y.-C. Zhang,
                  Phys. Rep. {\bf 254}, 215 (1995).
\bibitem{Kardar}  M. Kardar, Nucl. Phys. B {\bf 290} [FS {\bf 20}],
                  582 (1987).
\bibitem{MK}      E. Medina and M. Kardar,
                  J. Stat. Phys. {\bf 71}, 967 (1993).
\bibitem{DIGKB}   V. S. Dotsenko, L. B. Ioffe, V. B. Geshkenbein,
                  S. E. Korshunov and G. Blatter,
                  Phys. Rev. Lett. {\bf 100}, 050601 (2008).
\bibitem{Zhang}   Y.-C. Zhang, Europhys. Lett. {\bf 9}, 113 (1989).
\bibitem{Zhang90} Y.-C. Zhang, Phys. Rev. B {\bf 42}, 4897 (1990).
\bibitem{Kolom}   E. B. Kolomeisky, Phys. Rev. B {\bf 45}, 7094 (1992).
\bibitem{KK07}    I. V. Kolokolov and S. E. Korshunov,
                  Phys. Rev. B {\bf 75}, 140201(R) (2007).
\bibitem{KK08}    I. V. Kolokolov and S. E. Korshunov,
                  Phys. Rev. B {\bf 78}, 024206  (2008).
\bibitem{PS}
                  M. Pr\"{a}hofer and H. Spohn,
                  Phys. Rev. Lett. {\bf 84}, 4882 (2000).
\bibitem{HL}      B. I. Halperin and M. Lax,
                  Phys. Rev. {\bf 148}, 722 (1966).
\bibitem{ZL}      J. Zittartz and J. S. Langer,
                  Phys. Rev. {\bf 148}, 741 (1966).
\bibitem{L67}     I. M. Lifshitz,
                  Zh. Eksp. Teor. Fiz. {\bf 53}, 743 (1967)
                  [Sov. Phys. - JETP {\bf 26}, 462 (1968)].
\bibitem{GM}      V. Gurarie and A. Migdal,
                  Phys. Rev. E {\bf 54}, 4908 (1996).
\bibitem{BFKL}    E. Balkovsky, G. Falkovich, I. Kolokolov and V. Lebedev,
                  Phys. Rev. Lett. {\bf 78}, 1452 (1997);
                  Int. J. Mod. Phys. B {\bf 11}, 3223 (1997).

\bibitem{FKLM}    G. Falkovich, I. Kolokolov, V. Lebedev and A. Migdal,
                  Phys. Rev. E {\bf 54}, 4896 (1996).
\bibitem{MSR}     P. C. Martin, E. Siggia and H. Rose,
                  Phys. Rev. A {\bf 8}, 423 (1973).
\bibitem{dD}      C. de Dominicis,
                  J. Phys. (Paris) {\bf 37}, c01-247 (1976).
\bibitem{Jans}    H. Janssen, Z. Phys. B {\bf 23}, 377 (1976).
\bibitem{Fogedb}  H. C. Fogedby, Phys. Rev. E {\bf 59}, 5065 (1999);
                  Physica A {\bf 314}, 182 (2002).
\bibitem{cond}    In more detail, the origin of this condition is
                  explained in the last paragraph of Sec. \ref{FreeIC}
                  and in Ref. \onlinecite{KK08}.
\bibitem{exp}     The same values of exponents follow also from
                  the generalization
                  of the Imry-Ma scaling argument constructed in Ref.
                  \onlinecite{MG} and based on the disorder-dependent
                  Gaussian variational approach of Ref. \onlinecite{GO}.
\bibitem{SVBO}    U. Schulz, J. Villain, E. Br\'{e}zin and H. Orland,
                  J. Stat. Phys. {\bf 51}, 1 (1988).
\bibitem{ill-reg} Some other manifestations of the same phenomenon
                  are discussed in Sec. IVB of Ref. \onlinecite{KK08}.
\bibitem{BD}      E. Brunet and B. Derrida,
                  Phys. Rev. E {\bf 61}, 6789 (2000).
\bibitem{MG}      C. Monthus and T. Garel,
                  Phys. Rev. E {\bf 69}, 061112 (2004).
\bibitem{GO}      T. Garel and H. Orland,
                  Phys. Rev. B {\bf 55}, 226 (1997).

\end{thebibliography}
\end{document}